\documentclass[aps,pra,twocolumn,showpacs,superscriptaddress,groupedaddress]{revtex4}  
\usepackage{graphicx}  
\usepackage{dcolumn}   
\usepackage{bm}        
\usepackage{amssymb}   
\usepackage{color}
\usepackage{amsmath}

\def \lvec{(\kern-.26em(}

\begin{document}

\title{Collision Kernels from Velocity-Selective Optical Pumping with Magnetic Depolarization}

\author{T. Bhamre} 
\affiliation{Department of Physics, Princeton University, New Jersey 08544, USA}
\author{R. Marsland III}
\affiliation{Department of Physics, Massachusetts Institute of Technology, Massachusetts 02139, USA}

\author{I. K. Kominis} 
\affiliation{Department of Physics, University of Crete, Heraklion 71103 Greece}
\author{B. H. McGuyer}
\altaffiliation{Present address: Department of Physics, Columbia University, 538 West 120th Street, New York, NY 10027-5255, USA.} 
\author{W. Happer} 

 \affiliation{Department of Physics, Princeton University, New Jersey 08544, USA}

\vskip 0.25cm

\date{\today}
\begin{abstract}
We experimentally demonstrate how magnetic depolarization of velocity-selective optical pumping can be used to single out the collisional cusp kernel best describing spin- and velocity relaxing collisions between potassium atoms and low pressure helium. The range of pressures and transverse fields used simulate the novel optical pumping regime pertinent to sodium guidestars employed in adaptive optics. We measure the precession of spin-velocity modes under the application of transverse magnetic fields, simulating the natural configuration of mesospheric sodium optical pumping in the geomagnetic field. We also provide a full theoretical account of the experimental data using the recently developed cusp kernels, which realistically quantify velocity damping collisions in this novel optical pumping regime. A single cusp kernel with a sharpness $s=13\pm 2$ provides a global fit to the K-He data.
\end{abstract}
\maketitle

We demonstrate a new way to measure the transition matrix $W(v,v')$ for velocity-changing collisions between ground-state alkali metal atoms and arbitrary buffer gas molecules in the classical limit. This matrix contains all the information required for computing the contribution of collisions to the evolution of non-equilibrium velocity distributions in the vapor, which is given in general by
\begin{align}
\frac{\partial p(v,t)}{\partial t}=-\gamma_{\rm vd}p(v,t)+\gamma_{\rm vd}\int dv' W(v,v') p(v',t)
\end{align}
where $p(v,t)$ is the time-dependent velocity distribution and $\gamma_{\rm vd}$ the rate of velocity-changing collisions. Precise measurements of this matrix have long been of interest as a way to place empirical constraints on the interatomic potential for alkali metal - buffer gas pairs \cite{GibbleCooper1991}. Two new applications demand an increasingly precise knowledge of $W(v,v')$. First of all, the ease with which non-equilibrium velocity distributions can be created through velocity selective optical pumping makes alkali-metal vapor cells an ideal candidate for investigating new theorems of non-equilibrium statistical mechanics \cite{Verley2012}. $W(v,v')$ is the transition probability that arises in Markov derivations of these theorems, and quantitative knowledge of its behavior for real systems would make such experiments easier to design and analyze. Secondly, laser guide stars employed in the adaptive optics systems of modern terrestrial observatories produce non-equilibrium velocity distributions of optically pumped mesospheric sodium atoms. The numerical models used for optimizing the laser parameters for maximum backscatter require $W(v,v')$ as input, but the dearth of empirical measurements for sodium-air collisions has forced current models to use a highly simplified model, where $W(v,v')$ is the Boltzmann distribution in $v$ independent of $v'$ \cite{Rochester2012}. 
\section{Theory}
The theoretical framework we employed to design the experiment and analyze our data is laid out in detail in \cite{Marsland12}. Here we summarize the points that are most directly relevant to the current work. 

We model the vapor cell as an ensemble of single atoms whose velocities are perturbed by collisions with molecules of an inert buffer gas and with the cell walls. Neglecting radiation-pressure effects relevant to laser cooling experiments, the equilibrium velocity distribution of atoms in the buffer gas is the Maxwell-Boltzmann distribution. It will be convenient to measure atomic velocities in units of the most probable speed $v_D = \sqrt{2k_BT/M}$ where $T$ is the temperature of the cell and $M$ is the mass of the atom. We will denote the projection $v$ of the atomic velocity onto the laser beam direction by $x=v/v_{D}$. Although the overall velocity distribution remains constant, the incident laser beam introduces correlations between the spins and the velocities: due to the Doppler shift, a laser at frequency $\omega$ excites a resonance at frequency $\omega_0$ only in atoms with velocity $x_0=(\omega-\omega_{0})/kv_{D}$ along the beam. This is the origin of `velocity-selective optical pumping' (VSOP), whereby the distribution of atomic spin states becomes velocity-dependent. We assume the laser is weak enough that the small fraction of atoms in the excited states at any given time can be ignored, so the overall effect of the laser is to modify the distribution of ground-state sublevels at the resonant velocities. The distribution among energy sublevels $|\mu\rangle$ for atoms with velocities between $x$ and $x+dx$ is described by the $(2I+1)(2S+1)\times (2I+1)(2S+1)$ density matrix $d\rho(x) dx=\sum_{\mu\nu} \chi_{\mu\nu}(x)|\mu\rangle\langle \nu|\, dx$, where $S$ and $I$ are the electron and nuclear spins, respectively, and $|\mu\rangle$ and $|\nu\rangle$ are eigenstates of the ground-state Hamiltonian. The latter includes hyperfine splitting and Zeeman coupling to an external magnetic field and the corresponding eigenvalues are $E_\mu$ and $E_\nu$.
\begin{figure}
\centering
\includegraphics[width=8.5 cm]{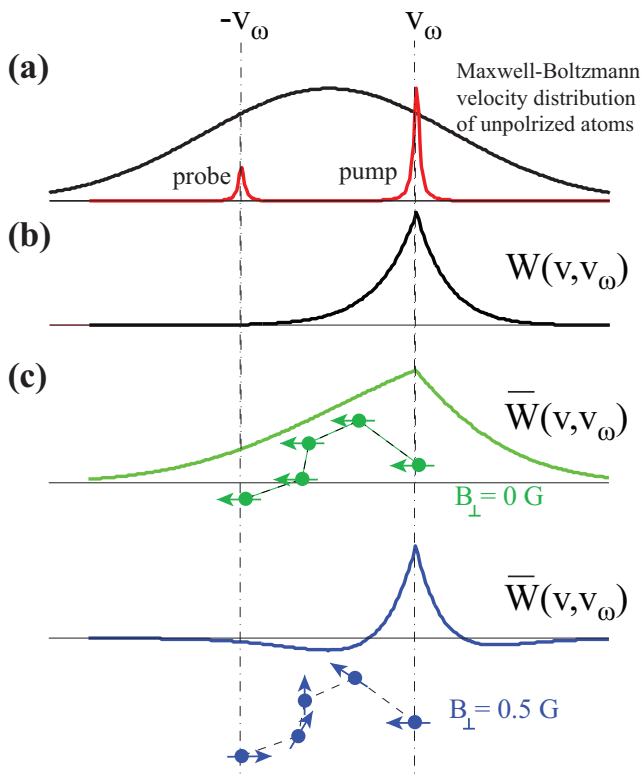}
\caption{\label{phys}(Color online)  (a) The monochromatic, circularly polarized pump laser of frequency $\omega$ selectively spin-polarizes atoms in the velocity group around $v_{\omega}$ that is Doppler-shifted into resonance. The counter-propagating probe beam of the same laser probes atoms at the velocity group around $-v_{\omega}$. Pumped atoms collide with buffer gas atoms and diffuse in velocity space from the velocity $v'$ before to the velocity $v$ after the collision. The velocity $v$ follows the probability distribution $W(v,v')$, shown in (b) for $v'=v_{\omega}$. By scanning the laser frequency $\omega$ the probe beam measures $W(-v_{\omega},v_{\omega})$. A transverse magnetic field causes the spins to precess during this process, acting as a "clock" during the velocity-space diffusion in between pumping and probing. (c) The overall effect of all the collisions is to produce a steady-state distribution of average spin as a function of velocity given by $\overline{W}(v,v_\omega)$ for two different values of the transverse magnetic field $B_{\perp}$. The effect of spin precession is seen in the sign reversal of $W(-v_{\omega},v_{\omega})$ at $B_{\perp}=0.5~{\rm G}$ relative to $W(-v_{\omega},v_{\omega})$ at $B_{\perp}=0~{\rm G}$.}
\end{figure}

In a VSOP situation with no velocity-changing collisions, some of these modes will have distributions $\chi_{\mu\nu}(x)$ sharply peaked around the resonant velocities $x_0$. As illustrated in Fig. \ref{phys}, adding an inert buffer gas allows atoms pumped at the resonant velocities to jump to other, non-resonant velocities through collisions with the buffer gas molecules that change the atomic velocity while preserving spin polarization. Fig. \ref{phys} also displays the effect of a transverse magnetic field, namely the spin will undergo Larmor precession while the atom diffuses from the pump resonant velocity to the probe resonant velocity, thus adding a new tunable timescale to the problem. 

The velocity-changing collisions are characterized by the `collision kernel' $W(x,x')$, which gives the probability that an atom with initial velocity $x'$ will have velocity $x$ after a single collision. Since our atoms remain in the spherically symmetric S state of orbital angular momentum for the overwhelming majority of the time, the collision kernel does not bear any dependence on the internal atomic state. The overall effect of collisions is to give $\chi_{\mu\nu}(x)$ a broad pedestal around the sharp resonant spike, and the shape of this pedestal contains valuable information about the collision kernel. To extract this information, we need to first write the steady-state solutions to the evolution equation for the $\chi_{\mu\nu}$'s in terms of $W(x,x')$, and then find a good ansatz form for $W(x,x')$ that can reproduce experimental data through appropriate adjustment of a few free parameters. 

The evolution of the velocity distributions $\chi_{\mu\nu}(x)$ of the spin modes is described by
\begin{equation}
\frac{\partial\chi_{\mu\nu}(x,t)}{\partial t}=-\int K_{\mu\nu}(x,x')\chi_{\mu\nu}(x',t) dx'+P_{\mu\nu}(x).
\label{ck2}
\end{equation}
where the damping kernel is
\begin{equation}
K_{\mu\nu}(x,x')=[\gamma_{\rm w}+i\Omega_{\mu\nu}+\gamma_{\rm vd}]\delta(x-x')-\gamma_{\rm vd} W(x,x').
\label{ck4}
\end{equation}
The rate of loss of polarized atoms due to diffusion or free flight to the walls is
$\gamma_{\rm w}$.  The Bohr frequencies $\Omega_{\mu\nu}=(E_{\mu}-E_{\nu})/\hbar$ depend on the applied magnetic field. For laboratory experiments with alkali-metal atoms in low buffer-gas pressures, the direct damping of spin modes by gas-phase collisions can be neglected, since practical buffer gases like N$_2$ or the noble gases He, Ne, Ar, {\it etc.} have such small spin-flip rates compared to $\gamma_{\rm w}$. The main effect of the buffer gas on spin relaxation is to lengthen the time needed for a polarized atom to diffuse to the wall, that is, to decrease $\gamma_{\rm w}$.  The rate of velocity relaxation due to the buffer gas is parametrized by the rate $\gamma_{\rm vd}$ at which atoms with initial velocity $x'$ are transferred to a range of final velocities $x$.  The distribution of final velocities is given by the collision kernel $W(x,x')$, which is the function we are trying to measure.

The laser-atom interaction that creates the non-equilibrium distribution of optically pumped atoms enters the model through the source term $P_{\mu\nu}=P_{\mu\nu}(x,\chi_{\mu'\nu'})$, which depends on the values $\chi_{\mu'\nu'}(x)$ of all the spin mode amplitudes at velocity $x$. For a true monochromatic light source exciting electrons to states with infinite lifetimes, $P_{\mu\nu}$ would be a superposition of delta functions. In fact, the finite excited-state lifetime, finite laser linewidth and other experimental issues to be discussed later combine to produce a Lorentzian source term with a particular width, which can be empirically determined from a measurement without buffer gas.

The nonequilibrium steady-state velocity distribution of the spin modes, found by setting $\partial \chi_{\mu\nu}/\partial t =0$ in Eq. (\ref{ck2}), is formally given by
\begin{equation}
\chi_{\mu\nu}(x)=\int K_{\mu\nu}^{-1}(x,x')P_{\mu\nu}(x')dx',
\label{ck6}
\end{equation}
where the inverse damping kernel is defined by $\int K_{\mu\nu}^{-1}(x,x')K_{\mu\nu}(x',x'')dx'=\delta(x-x'')$. Note that $P_{\mu\nu}$ depends on the $\chi_{\mu\nu}$'s, so this is not a closed-form solution, but we can use it to obtain a power series solution in the pump laser intensity, as shown explicitly in Eqs. (134) through (136) of \cite{Marsland12}. Our numerical calculations use the first-order solution, given by Eqs. (137) through (140) of \cite{Marsland12}. 

We now need a physically reasonable and computationally tractable Ansatz for $W(x,x')$ in order to extract information about the collisions from experimental signals. McGuyer \emph{et al.} have shown that `cusp kernels' $C_s(x,x')$ meet these requirements \cite{McGuyer12}. The cusp kernel function has a single free parameter $s$ called the `sharpness,' which we will employ as a fit parameter to match our model predictions to our experimental data. Setting $W(x,x')=C_s(x,x')$, the inverse damping kernel $K_{\mu\nu}^{-1}$ takes on the particularly simple form \cite{McGuyer12}:
\begin{align}
K&_{\mu\nu}^{-1}(x,x')=\frac{1}{\gamma_{\rm w}+i\Omega_{\mu\nu}+\gamma_{\rm vd}}\nonumber \\
&\times\left[\delta(x-x')+\frac{\gamma_{\rm vd}}{\gamma_{\rm w}+i\Omega_{\mu\nu}}C_{r_{\mu\nu}}(x,x')\right ],
\label{ck8}
\end{align}
where
\begin{equation}
r_{\mu\nu}=\frac{\gamma_{\rm w}+i\Omega_{\mu\nu}}{\gamma_{\rm w }+i\Omega_{\mu\nu} +\gamma_{\rm vd}}s.
\label{ck10}
\end{equation}

Using this result and Eq. (\ref{ck6}), we can obtain the steady-state $\chi_{\mu\nu}$'s in terms of the atomic parameters, $\gamma_{\rm w}$, $\gamma_{\rm vd}$ and $s$. The rates $\gamma_{\rm w}$ and $\gamma_{\rm vd}$ can be determined experimentally, so the solution has $s$ as its only free parameter. Once we calculate the probe beam absorption spectrum from the $\chi_{\mu\nu}$'s, we can easily find the sharpness $s$ that gives a least-squares best fit to the data. 

The experiment we performed is similar to the experiment of Aminoff \emph{et al.} \cite{Aminoff83}, the main difference being the addition of a transverse magnetic field introducing the magnetic depolarization discussed previously. The authors in \cite{Aminoff83} noted that the traditionally used Keilson-Storer kernel was a poor approximation of the actual kernel, since it could not fit the data over a broad range of pressures and the kernel's parameter was thus not unique but pressure-dependent. In contrast, we will demonstrate that a cusp kernel with a single sharpness can provide a global fit to the data. The additional, experimentally controllable parameter of the transverse magnetic field, incorporated in our formalism through the frequencies $\Omega_{\mu\nu}$, provides a systematic check of our understanding of the kernel, as it introduces an additional degree of freedom in the spin-velocity time evolution, not present in the previous experiments using longitudinal magnetic fields only.
\section{Experiment and Data Analysis}
\begin{figure}
\includegraphics[width=8.5 cm]{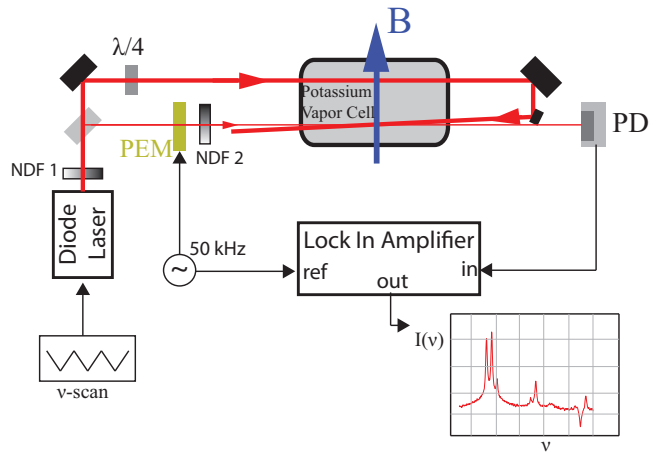}
\caption{(Color Online) Counter-propagating pump-probe lasers create and probe spin-velocity distributions of potassium metal in helium buffer gas. The atoms are contained in sealed glass cells of 1/2 in. diameter and 5/4 in. length, residing in an oven air-heated at $69\pm0.2~^{\circ}{\rm C}$ to maintain a high enough potassium vapor pressure of natural isotopic abundance, $93.26\%$ $^{39}$K and $6.73\%$ $^{41}$K. The cell and oven reside in the middle of  three pairs of Helmholtz coils along three mutually orthogonal axes providing a homogeneous, constant and tunable magnetic field. The intensities of the two counter-propagating beams were independently controlled with two neutral density filters. The beams are not perfectly antiparallel due to the pump mirror blocking the probe detection photodiode, but the angle between them was less than 5 mrad.}
\label{setup}
\end{figure}
In. Fig. \ref{setup} we depict the schematic of our experiment. We inserted three cylindrical sealed glass cells containing potassium and helium buffer gas at three different pressures into the path of two counter-propagating circularly polarized beams from a Toptica DL Pro tunable diode laser. The pump beam is responsible of VSOP, while we monitor the transmission of the weaker probe beam. 
To amplify the signal, we modulated the circular polarization of the probe beam with a photo-elastic modulator (PEM), cycling the polarization from right-circular to left-circular and back at a frequency of 42 kHz. Since both polarizations should have the same transmission efficiency with no optical pumping, we were able to extract and amplify the correction to the lineshape due to optical pumping using a lock-in amplifier phase-locked to the PEM, while slowly scanning the laser frequency over the potassium's Doppler profile.
\begin{figure*}[!]
\includegraphics[width=17 cm]{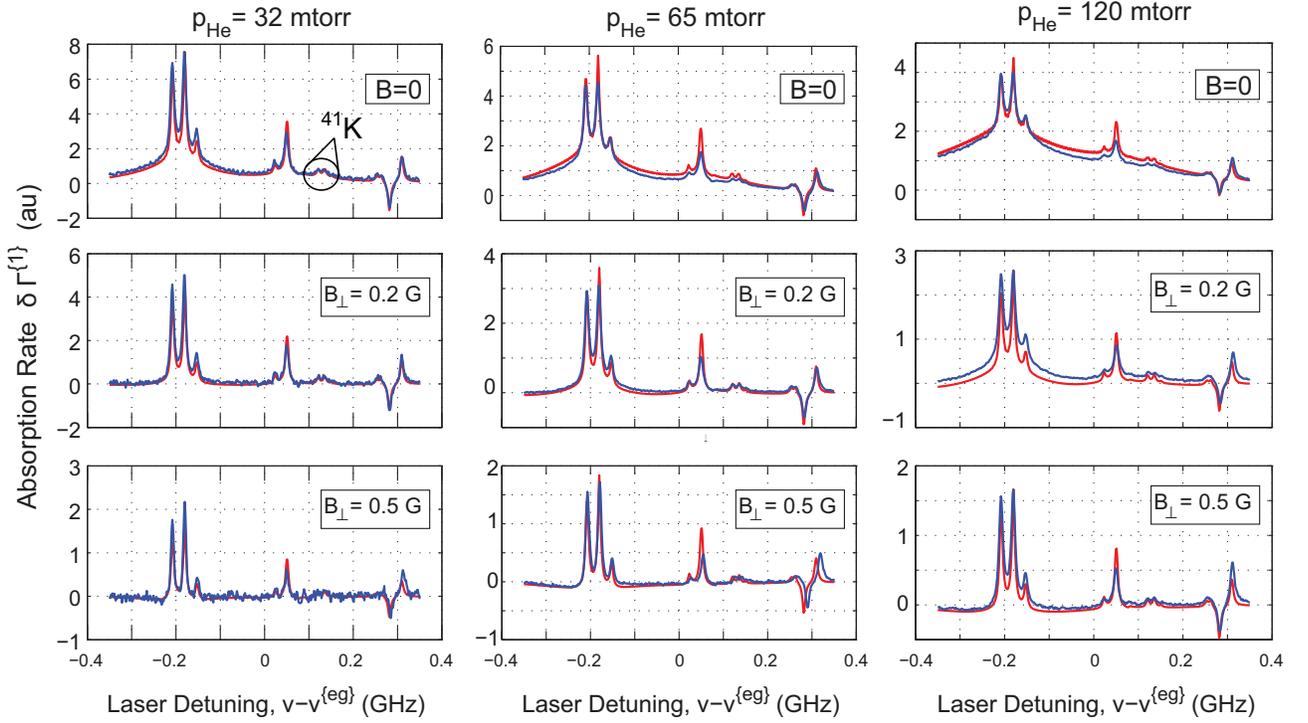}
\caption{(Color) Potassium-Helium data. Shown is the measured (red) and calculated (blue) absorption rate of the probe beam for three different magnetic fields (zero field, 0.2 G and 0.5 G, transverse to the laser propagation direction) at pressures of 32, 65, and 120 mTorr. The theoretical calculation was fit to the data by a least squares fit. A single cusp kernel with sharpness $s=13$ provides a good fit to all data. }
\label{data}
\end{figure*}

Since the pump beam only pumps atoms having velocities $x_{\mu\bar{\nu}}=(\omega-\omega_{\mu\bar{\nu}})/kv_D$, where $\omega_{\mu\bar{\nu}}$ is the frequency of the transition from ground state sublevel $\mu$ to excited state sublevel $\bar{\nu}$, and the probe beam only interacts with atoms with opposite velocities $-x_{\mu\bar{\nu}}$, the spectrum of probe beam transmission will display sharp saturated-absorption spikes at $\omega_{\mu\bar{\nu}}$ as well as the cross-over lines $(\omega_{\mu\bar{\nu}}+\omega_{\mu'\bar{\nu'}})/2$, where the pump and probe beams can simultaneously interact with the same atoms. To ensure that the probe beam does not modify the atomic density matrix, we measured the height of the tallest one of these spikes at various probe intensities and a constant pump intensity. At low probe intensities, the height scales linearly with intensity, while it becomes quadratic as the probe intensity is increased. By chosing a ratio of probe to pump intensity of about 10\%, we ensured we are well within the linear regime. Similarly, we can neglect the second and higher-order terms in pump beam intensity from the series solution of Eq. (\ref{ck6}) by examining the resonance's scaling with pump intensity and staying within the linear regime. As indicated in Fig. \ref{setup}, the filter (NDF1) for adjusting the pump intensity was positioned before the beamsplitter, so that we could adjust the pump intensity at constant probe/pump intensity ratio. In the first-order regime, the transmitted intensity should scale linearly with each beam intensity individually, and so their simultaneous variation should produce a quadratic shape. We set NDF1 so that we are well within this quadratic regime.

In Fig. \ref{data} we show the data acquired from our three cells at three different transverse magnetic fields for each cell, along with the theoretical fits. To compare the measured lineshapes to the cusp kernel predictions we used Eqs. (143) through (147) of \cite{Marsland12} to obtain the first-order lineshape corrections from the spin mode amplitudes $\chi_{\mu\nu}$. The required rates $\gamma_{\rm vd}$ and $\gamma_{\rm w}$ were obtained from the expressions
\begin{align}
\label{vd}
\gamma_{\rm vd}=\frac{v_D^2}{2D}(s+1)
\end{align}
and
\begin{align}
\frac{1}{\gamma_{\rm w}}=\frac{a}{v_D}+\frac{a^2}{8D}\left[ 1+4\ln \left(\frac{b}{a}\right)\right]
\end{align}
given in Eqs. (A1) and (79) of \cite{Marsland12} respectively. The most probable speed $v_D$ and the sharpness parameter $s$ have already been defined. The laser beam and cell radii are $a = 0.06$ cm and $b=0.56$ cm, respectively, and  $D$ is the diffusion coefficient of potassium atoms in helium, inversely proportional to the helium pressure. We used the value $D_0=0.45$ cm$^2$/s for 1 atm of helium. Note that Eq. (\ref{vd}) is valid if the collision kernel is adequately modeled by a single cusp kernel.

To correct for potassium's optical thickness, leading to pump and probe intensity variation along the length of the cell, we divide the data by the factor
\begin{align}
f(\omega)=\frac{e^{-\sigma(\omega) n l}}{\sigma(\omega)}(1-e^{-\sigma(\omega) n l})
\end{align}
where $\sigma(\omega)$ is the absorption cross-section of the unpolarized potassium atoms, $n$ is the number density of potassium atoms (a known function of temperature), and $l=2.7$ cm is the length of the cell. 

We also had to account for the fact that our pressure gauge had a large uncertainty at the low pressures we were using. To get a good fit, we had to adjust the pressure values input to the model within the range of uncertainty, and ultimately used the values 32, 65, and 120 mTorr instead of the nominal 25, 50, and 100.  Finally, we added one more free parameter to match the absolute scale of the y-axis in Fig. \ref{data} between model and data. We used the same scale factor for all nine data traces in Fig. \ref{data}. The simultaneous least-squares fit of the model to the nine data traces is superimposed on the data in Fig. \ref{data}, resulting in a sharpness of $s=13\pm 2$.

\section{Conclusions}
In conclusion, we have demonstrated a simple way to accurately determine the sharpness of cusp kernels modeling velocity-selective optical pumping. In particular, we have shown that a single cusp kernel is sufficient to reproduce lineshape measurements far from the Maxwell-Boltzmann equilibrium, across a variety of pressures and magnetic fields. The employed method of magnetic depolarization allows a rapid, experimentally straightforward and accurate measurement of collision kernels useful for constraining interatomic potentials, testing non-equilibrium statistical mechanics theorems, or optimizing laser guidestars systems.
\newline\newline
The authors are grateful to M. J. Souza for making cells, and to Ben Olsen, Natalie Kostinski, and Ivana Dimitrova for contributions to the experimental apparatus.  This work was supported by the Air Force Office of Scientific Research.

\end{document}